\documentclass[sigconf, balance, nonacm]{acmart}

\usepackage{graphicx}
\usepackage{subfiles}

\usepackage{todonotes}
\usepackage{booktabs}
\usepackage{array,multirow}
\PassOptionsToPackage{hyphens}{url}\usepackage{hyperref}

\usepackage{dsfont}
\usepackage{amsmath}

\usepackage{mathtools}

\usepackage{caption}
\usepackage{subcaption}

\usepackage{xcolor}

\usepackage{environ}

\NewEnviron{resizealign}{\sbox0{
    $\begin{matrix}\displaystyle\BODY\end{matrix}$}%
  \sbox1{$(\theequation)$}%
  \sbox2{\parbox{\dimexpr \wd0 + \wd1}%
    {\begin{align}\BODY\end{align}}}%
  \noindent\resizebox{\linewidth}{!}{\usebox2}%
}

\AtBeginDocument{%
  \providecommand\BibTeX{{%
    \normalfont B\kern-0.5em{\scshape i\kern-0.25em b}\kern-0.8em\TeX}}}

\copyrightyear{2024}
\acmYear{2024}
\acmConference[SIGIR '24]{Proceedings of the 47th International ACM SIGIR Conference on Research and Development in Information Retrieval}{July 14--18, 2024}{Washington D.C., USA}
\acmBooktitle{Proceedings of the 47th International ACM SIGIR Conference on Research and Development in Information Retrieval (SIGIR '24), July 14--18, 2024, Washington D.C., USA}

\begin{document}

\title{Efficiency-Effectiveness Tradeoff of Probabilistic Structured Queries for Cross-Language Information Retrieval}

\author{Eugene Yang}
\affiliation{%
  \institution{HLTCOE, Johns Hopkins University}
  \city{Baltimore}
  \state{Maryland}
  \country{USA}
}
\email{eugene.yang@jhu.edu}

\author{Suraj Nair}
\authornote{Work done prior to Suraj joined Amazon.}
\affiliation{%
  \institution{Amazon}
  \city{Seattle}
  \state{Washington}
  \country{USA}
}
\email{srnair@cs.umd.edu}

\author{Dawn Lawrie}
\affiliation{%
  \institution{HLTCOE, Johns Hopkins University}
  \city{Baltimore}
  \state{Maryland}
  \country{USA}
}
\email{lawrie@jhu.edu}

\author{James Mayfield}
\affiliation{%
  \institution{HLTCOE, Johns Hopkins University}
  \city{Baltimore}
  \state{Maryland}
  \country{USA}
}
\email{mayfield@jhu.edu}

\author{Douglas W. Oard}
\affiliation{
    \institution{University of Maryland}
    \city{College Park}
    \state{Maryland}
    \country{USA}
}
\email{oard@umd.edu}

\author{Kevin Duh}
\affiliation{%
  \institution{HLTCOE, Johns Hopkins University}
  \city{Baltimore}
  \state{Maryland}
  \country{USA}
}
\email{kevinduh@cs.jhu.edu}

\begin{abstract}
Probabilistic Structured Queries (PSQ) is a cross-language information retrieval (CLIR) method 
that uses translation probabilities statistically derived from aligned corpora.
PSQ is a strong baseline for efficient CLIR using sparse indexing.
It is, therefore, useful as the first stage in a cascaded neural CLIR system
whose second stage is more effective but too inefficient
to be used on its own to search a large text collection.
In this reproducibility study, we revisit PSQ by introducing an efficient Python implementation.
Unconstrained use of all translation probabilities that can be estimated from aligned parallel text
would in the limit assign a weight to every vocabulary term,
precluding use of an inverted index to serve queries efficiently. 
Thus, PSQ's effectiveness and efficiency both depend on how translation probabilities are pruned.
This paper presents experiments over a range of modern CLIR test collections
to demonstrate that achieving Pareto optimal PSQ effectiveness-efficiency tradeoffs benefits from multi-criteria pruning, which has not been fully explored in prior work.  
Our Python PSQ implementation is available on GitHub\footnote{\url{https://github.com/hltcoe/PSQ}} and unpruned translation tables are available on Huggingface Models.\footnote{\url{https://huggingface.co/hltcoe/psq_translation_tables}} 
\end{abstract}


\maketitle

\section{Introduction}

While recent developments in IR have focused on neural approaches with dense vectors produced by pretrained language models, 
such as DPR~\cite{dpr} and ColBERT~\cite{colbert},
studies have demonstrated that incorporating lexical retrieval models into the neural pipeline is still beneficial~\cite{wang2022inspection, leonhardt2022efficient, kulkarni2023lexically, nair2023blade}. 
Specifically for Cross-Language Information Retrieval (CLIR), \citet{nair2023blade} showed that combining rankings from sparse and dense retrieval models yields a better Pareto-optimal efficiency-effectiveness tradeoff. 
Among sparse models, Probabilistic Structured Query (PSQ)~\cite{darwish2003probabilistic, wang2012matching, xu2000cross, kraaij2003embedding} is extremely efficient compared to neural retrieval models. 
Therefore, combining any model with PSQ can potentially receive an effectiveness boost without increased latency, assuming a suitable parallel implementation.

PSQ is a statistical term translation approach
designed for use with sparse retrieval models such as BM25 or Hidden Markov Model (HMM)
to perform CLIR. 
It translates text from one language to another on a term-by-term basis,
which can be performed in conjunction with other retrieval text preprocessing, such as stopword removal or stemming, to maximize retrieval effectiveness.

\begin{figure}
    \centering
    \includegraphics[width=\columnwidth]{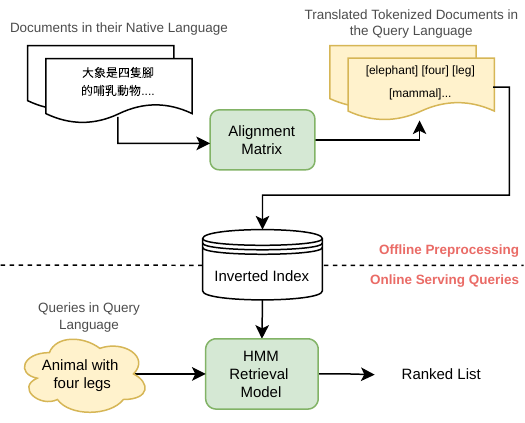}
    \caption{Indexing-time PSQ Retrieval Pipeline}
    \label{fig:pipeline}
\end{figure}

Because each term is treated independently, the translation mappings in PSQ can be computed either at query time or at indexing time. 
Most prior work on PSQ has focused on query-time implementation.
This limits effort at indexing time, but
at the cost of greater effort at query time,
where a few query words could expand
to hundreds of tokens~\cite{darwish2003probabilistic, kraaij2003embedding, wang2012matching}. 
While \citet{xu2000cross} proposed an indexing-time implementation of PSQ
that aims for better query-time efficiency by offloading most computation to indexing time, 
their implementation did not focus on pruning the very large number of possible translations.
Thus it was not compatible with an inverted index, limiting its practical use in large-scale applications. 
To further understand the tradeoff between retrieval effectiveness and efficiency, 
this work implements an efficient indexing-time PSQ.  
Figure~\ref{fig:pipeline} summarizes our process.

In statistical alignment models that estimate alignment probabilities based on token co-occurrence,
each term has a long tail of alternate translations. 
While including more translations improves retrieval effectiveness,
the resulting inverted index becomes dramatically larger, degrading efficiency. 

\citet{wang2012matching} experimented with three pruning strategies: 
Probability Mass Function (PMF) (a minimum threshold on the translation probability);
Cumulative Distribution Function (CDF) (a maximum threshold on the cumulative probability included in the alignment for a given source token);
and Top-k threshold (a maximum threshold on the number of alternative translations). 
The original paper studied the impact of applying each pruning strategy individually. 
Combinations of the three have been applied in practice to trim the alignment table~\cite{nair2023blade}, but a broad range of possible combinations has yet to be studied.

This reproducibility study focuses on revisiting PSQ and analyzing the efficiency and effectiveness tradeoff in a Pareto framework with multiple factors. 
We primarily focus on reproducing the findings from \citet{wang2012matching} because of its potential impact in modern IR algorithms, such as SPLADE~\cite{formal2021splade} and BLADE~\cite{nair2023blade}, that also exhibit an efficiency-effectiveness tradeoff in pruning the index. 
We revisit these pruning techniques with a modern technology stack, including more and cleaner parallel text, indexing-time processing, a public implementation that we will share, and larger evaluation collections.  
Specifically, we reimplement the indexing-time PSQ-HMM retrieval model introduced by \citet{xu2000cross}. 
To further understand the tradeoff, we conduct additional analysis on applying a combination of pruning techniques with Pareto optimality.

The contributions of this work are: 
(1) an efficient indexing-time PSQ implementation in Python;
(2) experimental results with modern, large CLIR evaluation collections and with more parallel text; and 
(3) a multi-factor analysis of the efficiency-effectiveness tradeoff for index pruning.
\section{Review of PSQ Development}

Early work on CLIR for free text
relied on comparable corpora~\cite{landauer1990statistical,sheridan1996experiments},
dictionary-based techniques~\cite{ballesteros1997phrasal},
or the use of rule-based Machine Translation (MT)~\cite{oard1998comparative}.
However, CLIR researchers soon began to use parallel corpora,
either directly~\cite{mccarley1999should} or through statistical MT~\cite{wu2008study,herbert2011combining}.
The basic approach to the direct application of parallel corpora was to first estimate translation probabilities from term alignments,
and then use those probabilities to weight the contribution of each translation
in the computation of the document scores on which ranking was based.

Early work with this technique used a smoothed unigram language model as the ranking function~\cite{xu2000cross},
although \citet{darwish2003probabilistic} later showed that the same approach could be used with other ranking functions.
Drawing insight from Structured Queries~\cite{pirkola1998effects}, a dictionary-based CLIR technique,
\citet{darwish2003probabilistic} gave the name Probabilistic Structured Queries (PSQ) to the general approach.
One way to conceptualize PSQ
is to think of the translation probability matrix as a lens
through which to view a document-language term frequency vector.
The result of this matrix-vector product is a vector of expected values for query-language term frequencies.
That vector can then be used with any ranking algorithm that uses term frequencies,
such as BM25 or unigram language models.
This same idea has also been used for annotation projection, in which labels such as syntactic structures \cite{hwa_resnik_weinberg_cabezas_kolak_2005}, semantic roles \cite{aminian-etal-2019-cross}, or sentiment \cite{rasooli18} are transferred from corpora in one language to another.

This broad idea still leaves a large design space.
\citet{darwish2003probabilistic} experimented with estimating Inverse Document Frequency (IDF) in a similar way,
since BM25 requires IDF estimates; \citet{montazeralghaem2019term} have subsequently shown that estimation can be improved.
However, \citet{nair2020experiments} found that estimating query-language IDF from a side collection
such as the New York Times\footnote{\url{https://catalog.ldc.upenn.edu/LDC2008T19}}
works well;
that is the approach we use in this paper.
Axiomatic analysis has recently suggested that the use of IDF,
as in BM25,
may also have theoretical advantages over unigram language models
that seek to achieve a similar effect using inverse collection frequency~\cite{rahimi2020axiomatic}.

Early PSQ experiments
(including its predecessors specific to unigram language models)
relied on unidirectional GIZA++\cite{och-ney-2003-systematic} alignments,
and subsequent techniques for improving alignments~\cite{denero2007tailoring,haghighi2008learning,dou2021word} have been found to further improve PSQ~\cite{wang2012matching}.

Probabilities estimated from parallel text are typically pruned,
both for practical reasons
(dense translation matrices would be very large)
and because small probabilities have small effects.
Alignment systems typically suppress probability estimates below some value
($10^{-6}$ in our work);
we refer to this as filtering on the \textit{Probability Mass Function (PMF)}.
\citet{wang2012matching} also experimented with retaining only enough alternate translations to reach some total probability;
we call this filtering on the \textit{Cumulative Distribution Function (CDF)}.
A third alternative is to retain only some number $k$ of the highest probability translations;
we call this \textit{top-k} filtering.
\citet{wang2012matching} found that each approach yielded similar results
by their Mean Average Precision (MAP) measure,
which rose as more translations were retained,
and then, beyond some point, fell.
Importantly, however, they renormalized after filtering;
whether to do so is another design decision.

The popular focus on a query-time implementation of PSQ is often justified by claiming that it
is more efficient than an indexing time implementation.
While that is certainly true for experimentation with typical test collections
that have many documents and few queries,
there are also applications (e.g., Web search)
in which the query volume can exceed even the volume of the indexed content.
Moreover, indexing-time implementations offer the potential for some elasticity
(e.g., servers can be allocated at lower-workload times);
queries, on the other hand, must be responded to in real-time as they arrive.
In this paper we therefore focus on indexing-time implementation of PSQ.

Neural CLIR methods have recently been shown to yield better rankings than PSQ~\cite{nair2022transfer}.
Due to the computational cost of neural methods,
they are frequently used in cascades in which sparse methods like PSQ generate initial results for neural reranking~\cite{nair2023blade} and in system combinations, since PSQ and neural CLIR have some complementary strengths.
PSQ thus lives on, even in a neural world.
\section{Generating Translation Probabilities}

Word alignments are needed in many applications, most notably in statistical Machine Translation (MT).
The seminal work of \citet{brown-etal-1993-mathematics} introduced IBM Models 1-5
to estimate word alignments from parallel corpora. 
Using Expectation-Maximization, these generative models infer latent word alignments from document-aligned pairs of texts in source and target languages.
IBM Model 1 computes the lexical translation table by aligning words independently.
Model 2 allows the alignment to be dependent on the respective sentence lengths. 
Model 3 introduces a fertility model, which indicates how many target words a source word may translate to. 
Model 4 accounts for reordering, modeling varying word orders across languages.
Like Model 4, Hidden Markov Model (HMM) alignment \cite{vogel-etal-1996-hmm} captures first-order dependencies,
where alignments depend on previously aligned tokens.\footnote{We note that HMM alignment differs from HMM ranking.  In HMM ranking (as used, for example, in PSQ-HMM) what has been called the query likelihood ranking model is implemented as an HMM.} 
GIZA++ \cite{och-ney-2003-systematic} is a toolkit that includes implementations of IBM Models 1-5 and the HMM model.

Using word alignments, we can generate lexical translation tables by aggregating the co-occurrence counts of aligned word pairs. 
GIZA++ performs alignments in two directions, source-to-target, and target-to-source. 
In the source-to-target direction, GIZA++ aligns words in the source language to words in the target language to estimate $P(target|source)$
and normalizes their sum to one. 
Similarly, in the target-to-source direction, GIZA++ aligns words in the source language to the words in the target language to estimate $P(source|target)$ and normalizes their sum to one.
\citet{wang2012matching} explored both possible normalization directions: the probability of a query-language term given a document-language term, or the probability of a document-language term given a query-language term.  They found the best results from the first of those, and that is therefore the direction in which we normalize for this paper.
Furthermore, we can concatenate the outputs of different word aligners to estimate a single lexical translation table, following \citet{zbib2019neural} and \citet{nair2020combining}.

Using such combined word alignments, the co-occurrence counts of word pairs that appear in multiple aligners increase, subsequently increasing the likelihood of these word pairs being translations~\cite{wu-wang-2005-improving}.
From here on, we refer to the lexical translation table produced by the word aligners as an alignment matrix.

\begin{table*}[t]
\caption{Collection Statistics. We use the same alignment matrix for NTCIR-8 and NeuCLIR Chinese.}
\label{tab:collection-stats}

\centering
\begin{tabular}{l|cccc|c|ccc}
\toprule
& \multicolumn{4}{c|}{CLEF 2003} & NTCIR-8 & \multicolumn{3}{c}{NeuCLIR 2022} \\
                          & French & Italian & German & Spanish & Chinese & Chinese & Persian & Russian \\
\midrule
\# of Parallel Sentences  &  17.6M &    3.6M &   4.4M &   15.7M &   12.1M &   12.1M &   20.8M &   14.5M \\
\midrule
\# of Docs                &   130K &   158K  &   295K &    454K &    309K &  3,179K &  2,232K &  4,628K \\
\# of Topics              &     60 &     60  &     60 &      60 &     100 &      49 &      46 &      45 \\
\# of Topics w/ Rel Docs  &     52 &     51  &     56 &      57 &     100 &      49 &      46 &      45 \\
\bottomrule
\end{tabular}

\end{table*}

\section{PSQ-HMM}

\newcommand{\doclang}{\ensuremath{\mathcal{S}}}
\newcommand{\querylang}{\ensuremath{\mathcal{T}}}
\newcommand{\bitext}{\ensuremath{B_{\doclang, \querylang}}}

In this section, we summarize the overall PSQ-HMM retrieval model pipeline,
which is an aggregation of the proposals by \citet{xu2000cross}, \citet{darwish2003probabilistic}, \citet{croft2010search}, and \citet{wang2012matching}. 

Based on \citet{xu2000cross} and \citet{darwish2003probabilistic},
we translate the documents $D_\doclang$ in language $\doclang$ into query language $\querylang$
with an alignment table trained on parallel corpus $\bitext$.
The probability that query language token $w_\querylang$ represents content in document $D_\doclang$
is expressed as

\begin{equation*}
    P(w_\querylang | D_\doclang) = 
    \sum_{\substack{w_\doclang \in D_\doclang}} 
    P_B(w_\querylang | w_\doclang) P(w_\doclang | D_\doclang)
    \nonumber
\end{equation*}

\noindent
where $P(w_\doclang | D_\doclang)$ is the probability of token $w_\doclang$ in document $D_\doclang$,
estimated from its term frequency;
and $P_B(w_\querylang | w_\doclang)$ is the alignment probability between tokens $w_\querylang$ and $w_\doclang$, estimated on $B$. 

Theoretically, every token in the query language has a non-zero probability of aligning with every token in the document language,
while only a few such alignments are meaningful.
Here, we estimate such probabilities through a parallel corpus $\bitext$ (denoted as $P_B$),
which limits the possible alignments of a given token $w_\doclang$ to the ones that co-occur in $\bitext$.
For simplicity, we denote the estimated probability of a token $w_\querylang$ being in document $D_\doclang$ as $P(w_\querylang | D_\doclang)$ without the $B$ subscript. 

\subsection{HMM Retrieval Model}

Assuming
a uniform prior on document relevance,
the probability that document $D_\doclang$ is relevant (denoted ``is R'') to query $Q_\querylang$ can be expressed as

\begin{align}
P(D_\doclang \text{ is R} | Q_\querylang) 
&= \frac{P(Q_\querylang | D_\doclang \text{ is R}) P(D_\doclang \text{ is R})}{P(Q_\querylang)} \nonumber \\
&\propto P(Q_\querylang | D_\doclang \text{ is R}) \nonumber
\end{align}

By modeling the relevance with an HMM~\cite{miller-hmm} model,
the overall relevance is the product of the probability of the document being relevant to each query term.
Therefore, the scoring function of a query-document pair integrated with PSQ is the probability $P(Q_\querylang | D_\doclang \text{ is R})$~\cite{xu2000cross},
which can be expanded as 

\noindent
\begin{resizealign}
&Score(Q_\querylang, D_\doclang) \nonumber 
= P(Q_\querylang | D_\doclang \text{ is R})  \nonumber \\
&= \prod_{w_\querylang \in Q_\querylang} \Bigl[
    \alpha P(w_\querylang | G_\querylang) 
    + (1-\alpha) P(w_\querylang | D_\doclang)
\Bigr] \nonumber  \\
&\propto \sum_{w_\querylang \in Q_\querylang} \log \Bigl[
    \alpha P(w_\querylang | G_\querylang) 
    + (1-\alpha) P(w_\querylang | D_\doclang)
\Bigr] 
\label{eq:log-hmm}
\end{resizealign}

\noindent
where $P(w_\querylang | G_\querylang)$ is the 
probability of $w_\querylang$ in a unigram language model
and $\alpha\in [0,1]$ is a Jelinek-Mercer smoothing hyperparameter to balance the two probabilities.
$P(w_\querylang | G_\querylang)$ can be estimated using an external corpus, as discussed in the next section.   As \citet{hiemstra2000using} has noted, what we call here HMM Retrieval is mathematically equivalent to a smoothed unigram language model, which has also been called the query likelihood model.

However, direct implementation of the smoothing in Equation~\ref{eq:log-hmm} would preclude using inverted index structures for efficient query processing because it gives a weight to every term in the vocabulary for every document.
In sparse retrieval models, such as BM25, 
the score of each document is the sum of its score for each query term.
In these models, the term score will be zero if the term does not appear in the document,
and those many zeros are what make inverted indexing efficient.   
To achieve that sparsity with an HMM retrieval model, 
we modify the similarity function by subtracting the baseline score of a document
with no overlapping query terms~\cite[Chapter~7.3.1]{croft2010search}.
This does not change the ordering of the documents but restores the sparsity.
Specifically, we define the scoring function as
\begin{resizealign}
Score(Q_\querylang, D_\doclang) 
& = \log P(Q_\querylang | D_\doclang \text{ is R}) \nonumber \\
& - \sum_{w_\querylang \in Q_\querylang} \log (\alpha P(w_\querylang | G_\querylang) ) \label{eq:hmm-subtract}
\end{resizealign}

Reorganizing Equation~\ref{eq:hmm-subtract},
the score of query term $w_\querylang$ for document $D_\doclang$ becomes zero
if the term does not appear in the document.
Therefore, we consider only overlapping terms between $Q_\querylang$ and the probabilistic translation of $D_\doclang$.
The scoring function becomes the sum of each term weight, which we define as $v_{w_\querylang}^{D_\doclang}$ for convenience,

\begin{resizealign}
& Score(Q_\querylang, D_\doclang) 
= \sum_{\substack{w_\querylang \in I_\querylang }} v_{w_\querylang}^{D_\doclang}  \nonumber \\
&= \sum_{w_\querylang \in I_\querylang} \log \left[
\frac{\alpha P(w_\querylang | G_\querylang) + (1-\alpha) P(w_\querylang | D_\doclang)}{\alpha P(w_\querylang | G_\querylang)} 
\right] \nonumber \\
&= \sum_{w_\querylang \in I_\querylang} \log \left[
\frac{(1-\alpha) P(w_\querylang | D_\doclang)}{\alpha P(w_\querylang | G_\querylang)} + 1
\right] \label{eq:sparse-score}
\end{resizealign}

\noindent
where $I_\querylang = Q_\querylang \cap \{ w_\querylang | P(w_\querylang | D_\doclang) > 0 \}$.
With Equation~\ref{eq:sparse-score},
the collection can be indexed in an inverted index
where the keys are query-language tokens.
At query time, we gather documents with at least one query term by traversing postings lists,
which is efficient with an inverted index.   

We implement the sparse retrieval with SciPy CSC sparse matrices\footnote{\url{https://docs.scipy.org/doc/scipy/reference/generated/scipy.sparse.csc_matrix.html}}
where columns represent the tokens in the query language and rows represent the documents. 
This is mathematically identical to using an inverted index,
since both approaches use the token as the lookup key for a list of documents that contain the token. 
It would be possible to further integrate our implementation with other sparse retrieval systems such as Lucene\footnote{\url{https://lucene.apache.org/core/}} and Pisa~\cite{mallia2019pisa}, which we leave for future work. 

At indexing time, the alignment matrix is transformed from dictionary form to a sparse matrix with pruning in preparation for fast indexing. Each document is tokenized and vectorized into a sparse vector and multiplied with the alignment matrix to get a sparse vector in the query language. 
The resulting vector is modified based on Equation~\ref{eq:sparse-score} to incorporate the query language unigram language model in its weights. 
Vectors are stored in chunks and then merged into a single inverted index, which is materialized by a CSC sparse matrix. 

At query time, the incoming query is preprocessed into a sparse vector
and multiplied by the inverted index sparse matrix.
Such multiplication is efficient and analogous to retrieving with an inverted index,
except for various posting list skipping techniques.

\subsection{Retrieval Efficiency}

At indexing time, each document is tokenized,
and its term frequency vector is then translated with the alignment matrix
as a dot product between the sparse vector in the document language $\doclang$
and the sparse alignment matrix $\doclang \rightarrow \querylang$.
This yields a sparse vector in the query language $\querylang$. 
For each translated term $w_\querylang$, we calculate its weight $v_{w_\querylang}^{D_\doclang}$ based on Equation~\ref{eq:sparse-score}. 
Documents are then indexed using the translated terms in the query language $\querylang$ as keys,
and each postings list identifying documents represented using the translated term, ordered by their weights $v_{w_\querylang}^{D_\doclang}$. 

Latency in serving the queries is strongly correlated with the length of the postings list for each query term under the same parallelism capability~\cite{macdonald2012learning}. %
By performing PSQ at indexing time, the number of alternate translations for each document term directly affects the size of the translated documents, and, thus, the length of each postings list. 
However, limiting the number of alternate translations also reduces retrieval effectiveness~\cite{wang2012matching}. 
This dilemma implies a clear tradeoff between retrieval effectiveness and efficiency. 

\section{Prior Findings on Pruning Translation Alternatives}

\citet{wang2012matching} explored three strategies for pruning the number of translation alternatives:
minimum translation probability (\textit{PMF Threshold});
maximum cumulative probability (\textit{CDF Threshold}); and
an integer threshold on the number of possible translations (\textit{Top-k} filtering). 

In their work, they evaluated the three strategies using query-side PSQ
on CLIR tasks of French and Chinese documents searched with English queries. 
\citet{wang2012matching} observed that applying CDF and PMF thresholds provides a broad range of efficiency
(measured in average number of translated query tokens)
and effectiveness (measured in percentage of monolingual MAP);
in their opinion, these were ideal choices for the efficiency-effectiveness tradeoff.
Top-k filtering, in contrast, requires more translation alternatives to perform reasonably well,
although it demonstrates a higher effectiveness peak. 

However, we argue that a broader range of efficiency and effectiveness values does not mean a better tradeoff. 
The quality of the efficiency-effectiveness tradeoff for a given pruning strategy should be measured by Pareto-optimality~\cite{nair2023blade}. One strategy with an higher Pareto frontier than another indicates being more effective with the same efficiency constraint.  
Therefore, we revisit the claim by comparing the three pruning strategies with a Pareto analysis framework. 

Additionally, there are a number of limitations in \citet{wang2012matching}: 
\begin{enumerate}
    \item they only evaluated with a query-side PSQ implementation;
    \item they did not consider combinations of pruning techniques; 
    \item they did not have access to as large a quantity of parallel text; and
    \item they did not have access to modern large CLIR test collections. 
\end{enumerate}
While influential work at the time, their conclusions may not hold  with a more efficient implementation and more rigorous evaluation.  
In the next section, we describe our modern experiment setup for PSQ.

\begin{table*}[t]
\newcommand{\q}{\textdaggerdbl}
\renewcommand{\d}{\textdagger}

\caption{Effectiveness, title queries, no index pruning. \q and \d indicate that PSQ-HMM is significantly different from QT-BM25 or DT-BM25, respectively, at $p<0.05$ by a two-tailed paired t-test.  Individual collection results have Bonferroni correction for 8 tests. Significance tests for the Average column are paired over the full set of 456 topics. }
    \label{tab:result-full-index}

    \centering

\newcommand{\STAB}[1]{\begin{tabular}{@{}c@{}}#1\end{tabular}}

\begin{tabular}{c|l|rrrr|r|rrr|r}
\toprule
& 
& \multicolumn{4}{c|}{CLEF 2003} 
& NTCIR-8  
& \multicolumn{3}{c|}{NeuCLIR 2022} 
& \multirow{2}{*}{Average} \\
&            & French & Italian& German & Spanish&Chinese & Chinese & Persian& Russian & {}\\
\midrule
\multirow{3.5}{*}{\STAB{\rotatebox[origin=c]{90}{R@100}}}
& QT-BM25    &  0.694 &  0.600 &  0.438 &  0.559 &  0.311 &   0.484 &  0.449 &  0.428 &  0.477 \\
& DT-BM25    &  0.735 &  0.673 &  0.629 &  0.567 &  0.441 &   0.479 &  0.475 &  0.436 &  0.546 \\
\cmidrule{2-11}
& PSQ-HMM    &  0.808 &  0.657 & \q0.677 &  0.620 &\q\d0.523 &   0.456 &  0.488 &  0.467 &\q\d0.585 \\
\midrule
\multirow{3.5}{*}{\STAB{\rotatebox[origin=c]{90}{MAP}}}
& QT-BM25    &  0.403 &  0.334 &  0.274 &  0.327 &  0.165 &   0.242 &  0.205 &  0.263 &  0.267 \\
& DT-BM25    &  0.381 &  0.348 &  0.376 &  0.336 &  0.262 &   0.264 &  0.209 &  0.245 &  0.302 \\
\cmidrule{2-11}
& PSQ-HMM    &  0.470 &  0.336 &  0.420 &  0.378 &\q0.318 &   0.238 &  0.212 &  0.257 &  \q\d0.332 \\

\bottomrule

\end{tabular}

\end{table*}

\begin{figure*}[t!]
    \centering
    \includegraphics[width=\textwidth]{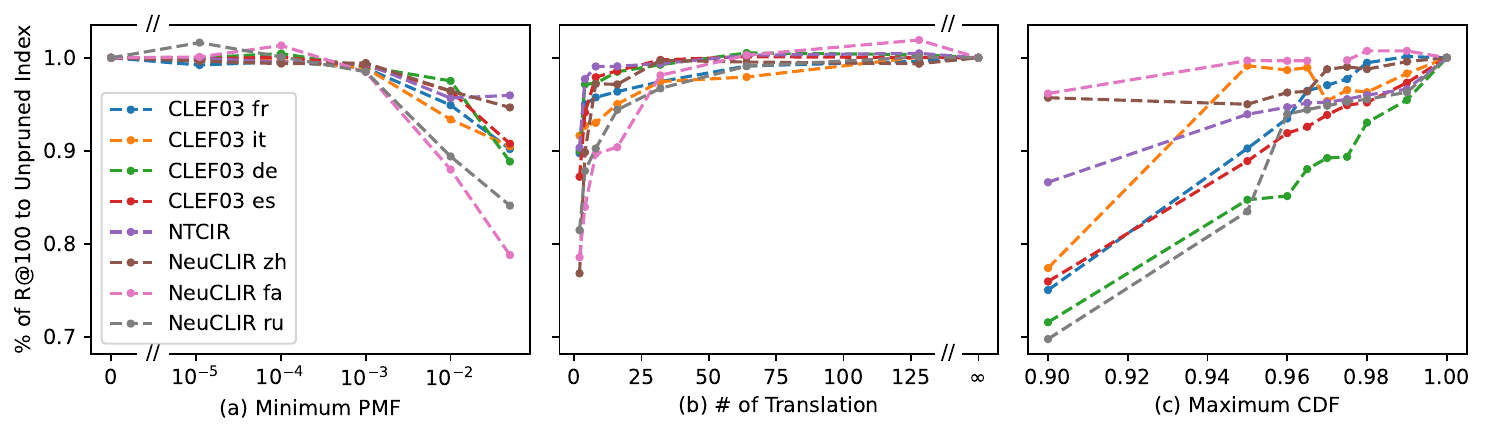}
    \caption{Relative R@100 score to the unpruned index of each collection. The x-axis of the PMF threshold graph is in log scale, and the unpruned index (0) is marked at the far left. }\label{fig:map-relative}
\end{figure*}

\section{Experiment Setup}

This section describes the design of our experiments.
We first describe the text alignment. 
Then we enumerate the hyperparameters used.
Next we discuss baselines, and finally we introduce our evaluation measures.

\subsection{Parallel Text Alignment}

To evaluate PSQ with modern resources, we train the alignment matrix with modern collections of parallel text, which have a far larger scale, and a  far greater diversity, than what was available to \citet{wang2012matching}. 
We use multiple sources from OPUS~\cite{tiedemann2012parallel},
including EuroParl~\cite{koehn2005europarl}, GlobalVoices,\footnote{\url{https://casmacat.eu/corpus/global-voices.html}} MultiUN~\cite{eisele-chen-2010-multiun}, NewsCommentary,\footnote{\url{https://data.statmt.org/news-commentary/v16/}} QED~\cite{abdelali-etal-2014-amara}, TED~\cite{reimers-2020-multilingual-sentence-bert}, and WMT-News~\cite{barrault-etal-2019-findings}. 
We also use bilingual Panlex dictionaries~\cite{kamholz-etal-2014-panlex} as an extra source of parallel text. 
The number of parallel sentences for each language is summarized in the top row of Table~\ref{tab:collection-stats}. 
To generate word alignments, we use three aligners: bidirectional GIZA++~\cite{och-ney-2003-systematic}, BerkeleyAligner~\cite{liang-etal-2006-alignment}, and Eflomal~\cite{ostling2016efficient}.
These individual word alignments are concatenated to create a unified alignment file.
Using this concatenated file, we proceed to create the unidirectional alignment matrix (i.e., the probability of a query-language (in our experiments, English) term given a document-language (non-English) term through maximum likelihood estimation.\footnote{\url{http://www2.statmt.org/moses/?n=FactoredTraining.GetLexicalTranslationTable}}
For bidirectional GIZA++, we run 5 iterations of Model 1 and HMM each, followed by 3 iterations of Model 3 and Model 4 each, using the Moses toolkit. 
The word alignments are estimated using the grow-diag-final-and heuristic from the bidirectional word alignments~\cite{koehn2007moses}.
For BerkeleyAligner, we obtain the word alignments by running 5 iterations of Model 1 followed by 5 iterations of HMM, both trained jointly in the bidirectional setting.
For Eflomal, we use the default settings and estimate word alignments using the grow-diag-final-and heuristic, similar to GIZA++.\footnote{\url{https://github.com/robertostling/eflomal/tree/7b97f19187c8b1bc1f21aefd77fc1b87575d1c00}}
We use the Google one-billon word corpus~\cite{chelba2013one} to estimate the query-language unigram language model.

\subsection{Evaluation Collections}

For retrieval evaluation, 
we use French, Italian, German, and Spanish collections from CLEF 2003~\cite{braschler2003clef},
Simplified Chinese from the NTCIR-8 ACLIA Task~\cite{mitamura2010overview},
and Chinese, Persian, and Russian from TREC NeuCLIR 2022~\cite{neuclir2022}. 
The NeuCLIR collections are much larger than those from CLEF and NTCIR.
Since our Chinese parallel text uses simplified characters,
we mapped traditional characters in the NeuCLIR Chinese collection to simplified characters
to match the alignment matrix.\footnote{\url{https://github.com/NeuCLIR/download-collection/blob/ad53164a90dbe0b1e6a9fc7e960e643f4cab7f4a/convert_chinese_char.py}} 
Collection statistics are summarized in Table~\ref{tab:collection-stats}. 
Our experiments use the English title field of the topics
(short, keyword descriptions of the information need)
as queries. 

For both parallel text and the information retrieval test collections,
we tokenize the text with Moses Tokenizer\footnote{\url{https://github.com/luismsgomes/mosestokenizer}} and remove punctuation, diacritics, stopwords (provided by NLTK~\cite{bird2009natural}), and casing
to normalize the text.
For Chinese, we segment sentences into words that may comprise more than one character using Jieba.\footnote{\url{https://github.com/fxsjy/jieba}} 
Matching text preprocessing across the two steps ensures consistency
when applying alignment matrices to retrieval documents, and in our experience failure to do so is one of the most common mistakes made by people who are new to PSQ.

\subsection{Hyperparameters}
For our main results, we create the inverted index without unnecessary pruning.
We simply impose an initial PMF threshold of $10^{-6}$
to ensure the resulting inverted index will fit in memory. 
To investigate the tradeoff between retrieval effectiveness and index size,
we experiment with three hyperparameters for trimming the alignment matrix:
minimum PMF, maximum CDF, and the $k$ for top-k filtering. 
All three approaches limit index size, but they operate differently. 
The experiment involves all combinations of six minimum PMF thresholds (0, 0.05, 0.01, 0.001, 0.0001, and 0.00001), eight values for top-k (2, 4, 8, 16, 32, 64, 128, and unrestricted) and ten CDF thresholds (0.800, 0.900, 0.950, 0.960, 0.965, 0.970, 0.975, 0.980, 0.990, and 1.000), resulting in 480 models per collection. 
We aim to cover values that one might consider using in a practical application, and also values that might catastrophically damage the retrieval results. 

We also experimented with other pruning methods
such as minimum document term weight $v_{W_\querylang}^{D_\doclang}$
and maximum number of translated tokens in a document.
We omit those results from this paper
as they generally offer a worse tradeoff between index size and retrieval effectiveness.

\subsection{Baselines}

We evaluate PSQ-HMM against BM25~($k_{1}=0.9, b=0.4$) with the use of one-best machine translation of either the query or the document, which are both strong CLIR baselines~\cite{neuclir2022, nair2022transfer}.
In prior work [anonymized], we had tried both HMM and BM25 retrieval models with one-best machine translation results, finding that the BM25 results were numerically slightly better. 
We therefore report results with BM25 as being representative of what can be achieved using one-best machine translation.
For query translation (QT), we translated the queries using Google Translate. 
For document translation (DT), we use public sets of translated documents provided in prior work~\cite{nair2022transfer, lawrie2023mlir},
which were generated by AWS Sockeye V2~\cite{sockeye2} trained on general domain parallel text. 

\subsection{Evaluation Measures}

We measure retrieval effectiveness using mean average precision (MAP) and recall at 100 (R@100),
which have been commonly reported in prior work on these test collections~\cite{nair2022transfer, nair2023blade, rahimi2020axiomatic, wang2012matching}.
We use R@100 to characterize retrieval effectiveness when used as the first stage of a neural reranking pipeline,
and MAP to characterize standalone use of a technique as a single-stage ranking model.
Following prior work, we only consider topics with at least one known relevant document when reporting effectiveness measures,
since neither of our measures can distinguish systems on topics for which no relevant documents are known to exist. 
For retrieval efficiency, we could use wall clock time to measure query latency,
but that could be affected by concurrent processes on the same machine.
To avoid that source of noise, we instead use index size as a proxy for query latency.
Query latency is linear in the length of the postings lists for the query terms,
and the index size is dominated by the postings lists
(the vocabulary is small enough to fit in memory).
Thus the total index size is a good proxy for query latency~\cite{macdonald2012learning}.

\section{Results and Analysis}

In this section, we first compare the effectiveness of PSQ-HMM on our test collections
with baselines using modern neural machine translation models,
which situates the performance of PSQ with respect to modern sparse CLIR models. 
We then revisit the claim of \citet{wang2012matching} by conducting a similar analysis,
applying each pruning technique individually. 
Finally, we present experiments applying a combination of techniques
to further analyze the efficiency-effectiveness tradeoff with Pareto optimality. 

\subsection{Effectiveness Compared to Baselines with Modern Machine Translation}

As summarized in Table~\ref{tab:result-full-index}, PSQ-HMM without index pruning
numerically outperforms both the use of Google for query translation (QT-BM25)
and the use of Sockeye for document translation (DT-BM25) in both MAP and R@100. 
Focusing first on R@100, without index pruning PSQ outperforms DT and QT on six of the eight collections
(the exceptions are Italian and NeuCLIR Chinese),
although the only statistically significant improvements are QT-BM25 for German and NTCIR Chinese and  DT-BM25 for NTCIR Chinese.
As Table~\ref{tab:collection-stats} shows, the number of topics in each individual collection is limited,
but across the eight collections, we have 456 topics.
That larger number is sufficient to detect statistically significant improvements over both QT-BM25 and DT-BM25
in the microaveraged R@100 across the eight collections,
weighting each collection in proportion to the number of topics that it contributes to the total.

\begin{figure*}[t]
    \centering
    
    \begin{subfigure}{\textwidth}
        \includegraphics[width=\textwidth]{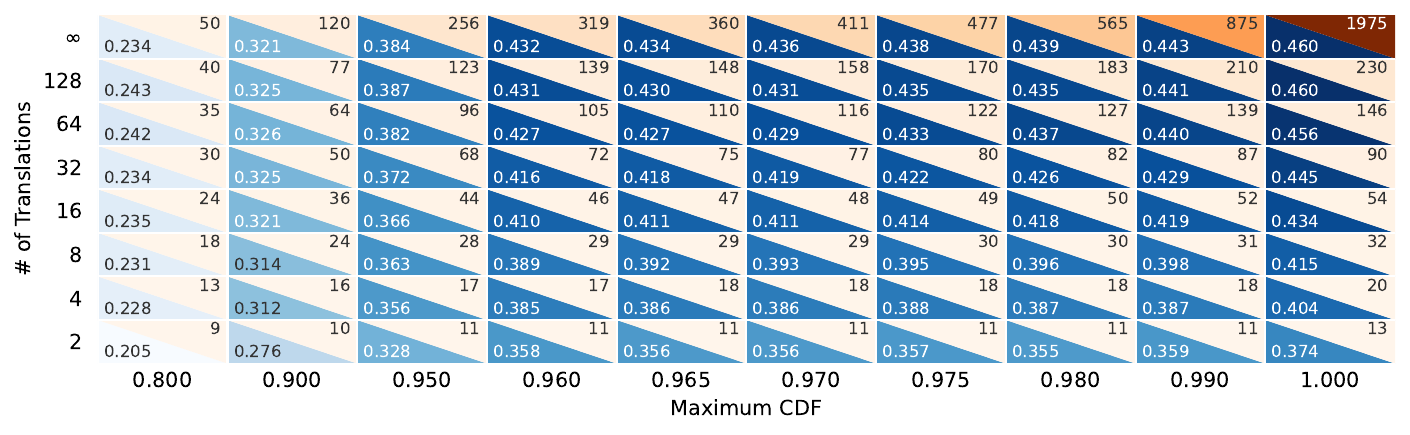}
        \caption{Top-k Filtering vs. CDF Thresholding}
    \end{subfigure}
    
    \begin{subfigure}{\textwidth}
        \includegraphics[width=\textwidth]{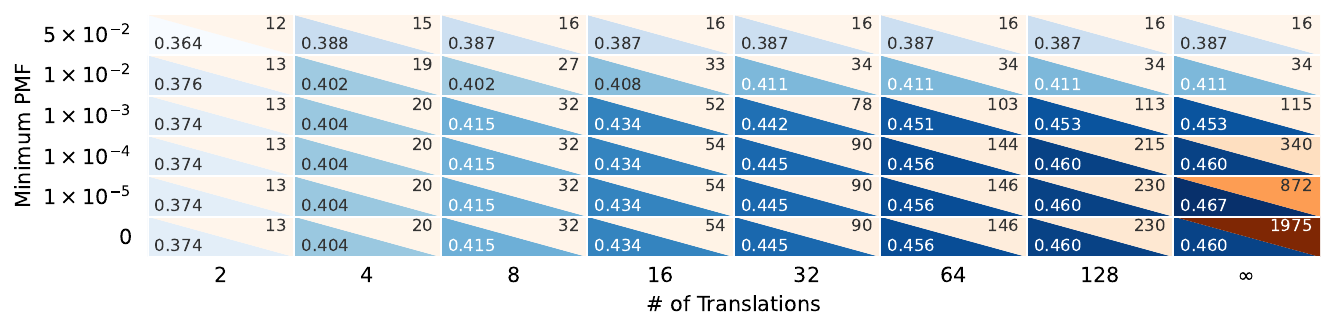}
        \caption{PMF Thresholding vs. Topk-k Filtering}
    \end{subfigure}
    
    \caption{R@100 and index size (GB) on NeuCLIR Russian. Blue and brown represent R@100 and index size, respectively; darker indicates larger values. The ideal is light brown (small index) and dark blue (high R@100). }
    \label{fig:heatmap-tradeoff}
    \vspace{-1em}
\end{figure*}

Looking now at the two languages for which we saw numerical decreases in R@100,
we focus first on NeuCLIR Chinese.
We transformed all traditional Chinese characters in that collection to simplified characters,
so character mismatch between the collection and the alignment matrix cannot explain the problem.
However, there are also vocabulary differences to be considered.
The vocabulary in the document collection, which includes a substantial number of documents from outside China, 
differs markedly from the vocabulary in the alignment matrix that was principally trained on parallel text from China.
Such a vocabulary mismatch could explain some systematic underperformance of PSQ on this collection.
Notably, for the NTCIR Chinese collection, which contains only documents from China,
PSQ statistically significantly outperforms both QT-BM25 and DT-BM25.  
For Italian, we note that it is only the DT-BM25 model that numerically outperformed PSQ-HMM.
Because this difference is not statistically significant,
this may simply be a random effect.  %

The MAP results at the bottom of Table~\ref{tab:result-full-index} show similar patterns.
Again, the microaverage over all 456 topics indicates that PSQ-HMM is statistically significantly better than either QT-BM25 or DT-BM25.
Again we see numerical underperformance of PSQ-HMM in Italian (below DT-BM25)
and NeuCLIR Chinese (below QT-BM25 and DT-BM25),
and we now also see NeuCLIR Russian slightly below QT-BM25.
Overall, we conclude that MAP and R@100 yield similar results when PSQ-HMM is used without index pruning.

We note that PSQ is far more efficient at indexing time than DT-BM25 when using a neural translation model,
since PSQ-HMM requires only a sparse table lookup to get translation probabilities and then a sparse matrix multiplication
to create sparse query-language vectors. 
At query time, all models use a sparse inverted index. 
However, PSQ-HMM indexes with an unpruned alignment matrix are far larger than an index built in the document language would be;
indeed, such an index can be as much as 20 times larger than the original document language text! 
Pruning the alignment matrix reduces not only index size but also query latency,
by limiting the number of documents with matching tokens.

\subsection{Efficiency-Effectiveness Tradeoff}

\begin{figure*}[t]
    \centering
    \includegraphics[width=\textwidth]{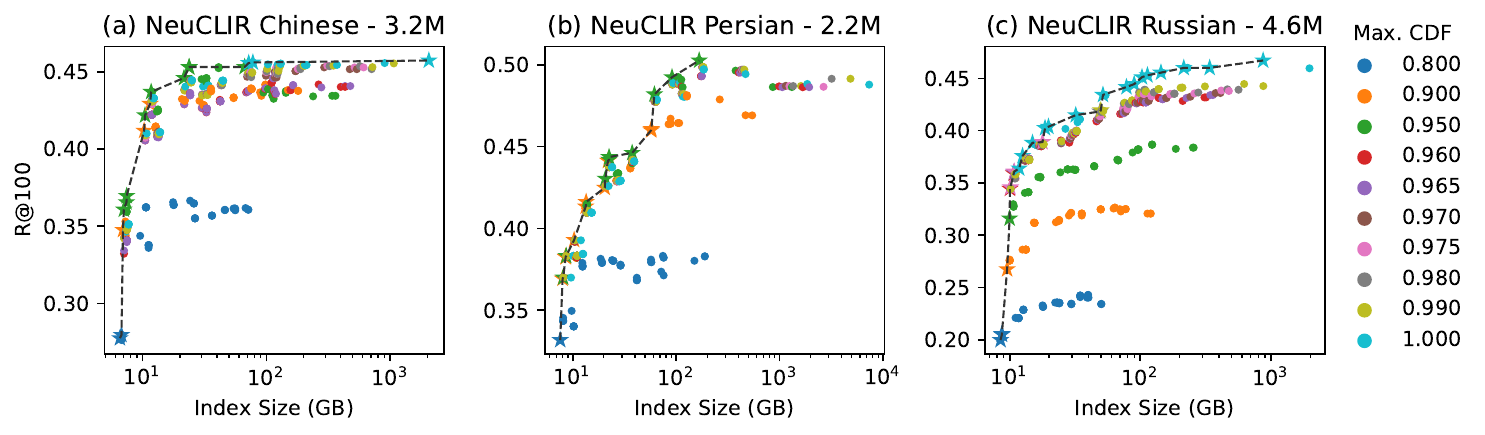}
    \includegraphics[width=\textwidth]{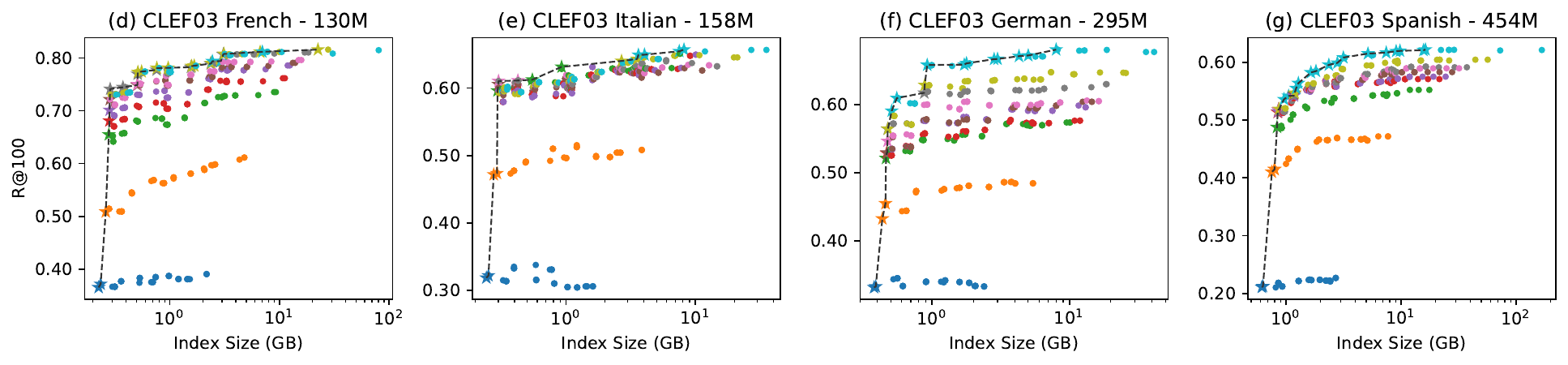}
    \caption{Pareto graphs. Stars indicate Pareto-optimal runs, i.e., those on the Pareto frontier. }
    \label{fig:pareto}
\end{figure*}

According to \citet{wang2012matching}, CDF and PMF thresholding result in a better tradeoff between efficiency and effectiveness compared to top-k filtering due to their larger range of possible outcomes. 
In this section, we revisit this finding
and include in the analysis combinations of pruning methods. 

As demonstrated in Figure~\ref{fig:map-relative},
holding everything else constant,
manipulating the maximum CDF has a larger effect on R@100 than using a PMF threshold or top-k filtering.  This differs from what \citet{wang2012matching} observed for MAP, the measure on which they focused. 
With a moderate degree of pruning (the middle parts of the graphs), both a PMF threshold and top-k filtering provide roughly 90\% of the R@100 of the full index.

However, similar to experiment results in \citet{wang2012matching},
retrieval results suffer when limiting top-k to a small number of translations (Figure~\ref{fig:map-relative}(b)). 
Smaller fan-out more closely approximates one-best translation,
which is essentially a statistical MT model without a target language model.
SMT generally underperforms neural MT by translation quality measures,
and those differences are reflected in retrieval effectiveness. 
However, using more translation alternatives offsets that,
achieving R@100 within 5\% of that of the full index. 
In this case, we observe very similar results for MAP. 

\citet{wang2012matching} conclude 
that a pruning strategy leading to a larger dynamic range conveys a preferable efficiency-effectiveness tradeoff. 
Our experiment results also suggest CDF thresholding has a larger range in R@100;
however, we conclude it actually introduces more risk of worse effectiveness. 
Thus, our analysis reveals that it is not preferable.

While pruning the alignment matrix can degrade retrieval effectiveness,
the resulting inverted index is smaller than an unpruned index, 
with different pruning techniques resulting in different drops in effectiveness.
Figure~\ref{fig:heatmap-tradeoff} shows that
different techniques also affect index size differently. 
At the upper right corner of Figure~\ref{fig:heatmap-tradeoff}(a),
where the number of translations is unlimited, and there is no thresholding on CDF,
PSQ provides the highest possible R@100 (0.460, in darkest blue),
but also the largest index (1,975 GB, in darkest brown).
We would prefer high effectiveness (dark blue) with a small index (light brown);
top-k filtering demonstrates this. 
In the rightmost column (using top-k filtering without a CDF threshold),
R@100 gradually drops with index size,
resulting, for example, in an index of size of 54GB and an R@100 of 0.434 on this Russian collection. 
Top-k filtering provides a better tradeoff between index size and effectiveness
than just a CDF threshold (top row),
which requires an index of 319GB to reach 0.432 R@100.
Applying both techniques (the cells in the middle),
always admits a setting without using a CDF threshold that achieves at least equal R@100 with a smaller index;
this indicates that CDF thresholding, contrary to conclusions in \citet{wang2012matching}, is less suitable for controlling 
the index size vs. R@100 tradeoff space.

Comparing PMF thresholding to top-k filtering,
both of which lead to smaller effectiveness drops than CDF thresholding in Figure~\ref{fig:map-relative},
top-k filtering is slightly better. 
With no PMF threshold (bottom row)
and allowing at most eight alternate translations,
we achieve 0.415 R@100 with an index size of 32 GB. 
On the other hand, with no limit on translation alternatives and a PMF threshold of 0.01,
we only get 0.411 R@100 with a larger 34 GB index. 
We observe a similar trend in MAP.

Other collections demonstrate similar trends to those seen on NeuCLIR Russian in Figure~\ref{fig:map-relative},
scaled according to the size of the collection. 
We picked the largest collection in our experiment as an example to demonstrate the effect more clearly. 

\subsection{Pareto Optimality}

\begin{figure*}[t!]
    \centering
    \includegraphics[width=\textwidth]{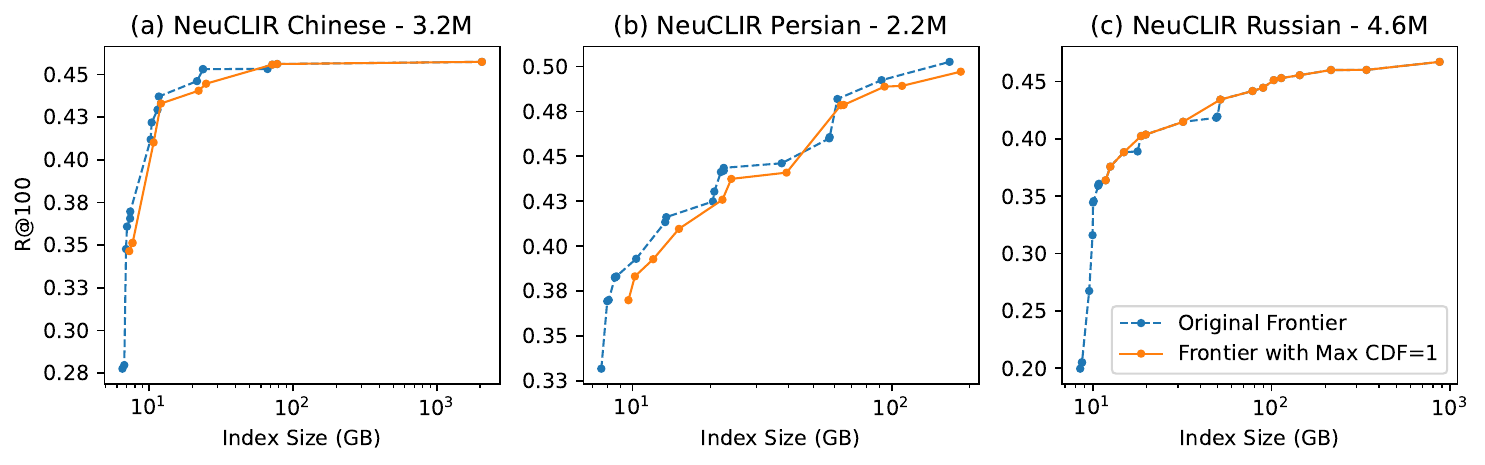}
    \includegraphics[width=\textwidth]{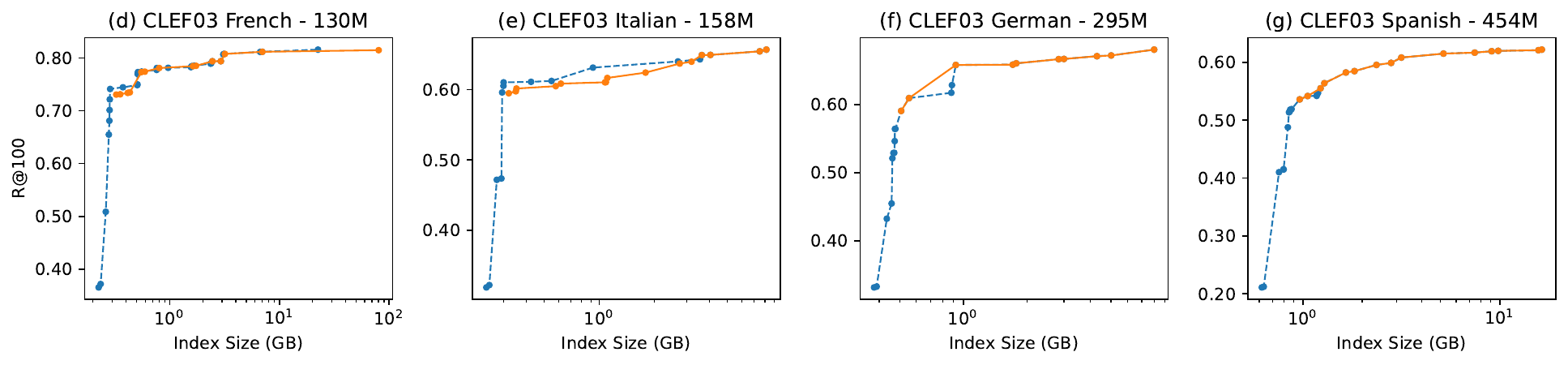}
    \caption{Comparison of Pareto Frontier between using and not using CDF thresholds.}
    \label{fig:pareto-compare-neuclir}
\end{figure*}

Finally, we plot all 480 pruning settings as a scatter plot for a Pareto analysis in Figure~\ref{fig:pareto}. 
Pareto-optimal settings
(those that achieve the highest R@100 without increasing index size)
are starred;
connecting them forms the Pareto frontier (dotted lines). 
On the frontier, all settings are Pareto optimal.
An implementer can then choose one setting over another by defining a utility function on that tradeoff.  

For all three NeuCLIR languages, there is a diminishing return with increased index size. 
With a large index,
R@100 can be lower than some settings with a smaller index (this is seen in Persian and Russian). 
That indicates a regularization effect,
stemming from a mild PMF or top-k filtering,
that trims noise from translations with very low probability. 
However, such effects are far less dramatic than the sharp falls observed in \citet{wang2012matching}. 

For the CLEF 2003 collection (Figure~\ref{fig:pareto}(d-g)), 
especially in German and Spanish, 
where the alignment matrices are well-trained,
the graphs demonstrate a fan-shaped pattern;
this indicates that there is also a diminishing return phenomenon 
among the 48 models with a given CDF threshold. 
For French, the graph demonstrates a small regularization benefit when imposing a light CDF threshold,
resulting in a Pareto improvement (i.e., pushing the Pareto frontier outward) over those without a CDF threshold.
However, indexes without a CDF threshold are still very close to the Pareto frontier,
which aligns with our observations and conclusions from the large NeuCLIR collections.

Most settings with no CDF threshold are on or near the Pareto frontier,
indicating that changing the index size by top-k filtering and a PMF threshold
suffices for Pareto optimality.
The left of each graph, where the index is small,
shows a steep drop in R@100;
such regions are unlikely to be used in practice
due to poor effectiveness. 
It is possible, and in many use cases preferable,
to achieve higher effectiveness with slightly larger indexes. 

Figure~\ref{fig:pareto-compare-neuclir} directly shows a comparison of the Pareto frontier between tuning with all three pruning techniques (480 models)
against using only a PMF threshold and top-k filtering, keeping the CDF threshold at 1.0 (48 models).
By the definition of Pareto optimality,
since the set of the models using all pruning techniques is a superset of a set that only uses two,
the Pareto frontier when using all techniques (blue dashed lines)
provides a tradeoff between index size and R@100 no worse than that using just two techniques.
However, applying only PMF and top-k filtering (orange solid lines)
is very close to, if not the same as, using all three. 
\section{Summary}

This paper has systematically studied indexing-time PSQ,
a strong sparse CLIR method aggregating the salient works by \citet{xu2000cross}, \citet{darwish2003probabilistic} and \citet{wang2012matching}. 
We revisit the three pruning techniques studied by \citet{wang2012matching}. %
Index pruning experiments and Pareto analysis
show that CDF thresholds provide a sub-optimal 
effectiveness-efficiency tradeoff,
but that PMF thresholds and top-k filtering
provide Pareto-optimal operating points evidenced by both R@100 and index size.

While PSQ was developed before the rise of neural IR models,
the effectiveness-efficiency tradeoff is similar when pruning models. 
Our findings and analytical methods can inform further tuning of neural IR models,
such as BLADE~\cite{nair2023blade},
which conceptually also transform documents into query language vectors at indexing time.

\bibliographystyle{ACM-Reference-Format}
\bibliography{custom}

\end{document}